\documentstyle[12pt,epsfig,a4]{article}
\input psfig.tex
\begin{document}
\noindent
{\tiny\bf To be published in the Proceedings  of
The  10th 
International
Summer School-Seminar on Recent Problems in
Theoretical $\&$ Mathematical Physics,} {\tiny\bf Kazan, Russia,
June 22 -July 3, 1998} \\

%\vspace*{2.5cm}
\begin{center}

{\bf ON THE FORMATION AND DYNAMICS OF MASS-OUTFLUX FROM MATTER ACCRETING ONTO COMPACT OBJECTS}\vspace{1.3cm}\\
\medskip

{\bf Tapas Kumar Das$^{1}$}
 \vspace{0.3cm}\\
 $^{1}$ Theoretical Astrophysics Group \\
 S. N. Bose National Centre For Basic Sciences \\
 Calcutta 700 091 India \\
 E-mail tdas@boson.bose.res.in
\end{center}

\vspace*{0.5cm}

\noindent
{\small Active galaxies and quasers are believed to harbour Black Holes at their centers
and at the same time produce cosmic radio jets through which immense
amount of matter and energy are ejected out of the core of the galaxy. In our
work we compute the mass outflow rates from advective accretion disks around black 
holes, which, in principal could explain the origin of jet formation
in self-consistent manner. These computations are done using combinations 
of exact transonic inflow and outflow solutions which may or may not form
shock waves. The approach of our calculation is somewhat different from
that used in the literature so far. Our work 
for the first time connects the accretion
type and wind type topologies self consistently.
Our result, in general, matches with numerical simulation works and successfully
points out to non-steady behavior which may evacuate the disk producing 
quiescene states.}\\[0.5cm]
\noindent
{\Large\bf Introduction }\\[0.25cm]
\noindent
It is now widely believed that active galaxies and quasers harbour compact objects at their centres.
One of the prominent signatures of activity around these objects is the generation of mass outflow
in the form of jets. AGNs produce cosmic radio jets through which immense amount of matter and energy
are spurted out of the core of the galaxies. Even the micro-quasers having stellar mass black holes
at their centres have recently been discovered which exibit same sort of phenomena. Sometime these
outflows show superluminal motion also. The existing models in the present literature which study
the formation and  dynamics of mass outflow are roughly of three types. The first type
of solutions confine themselves to the jet properties only, completely decoupled from the internal
properties of accretion disks [1]. In the second type, efforts are made to correlate the internal
structure of the accretion disks with that of the outflow using both hydrodynamics [2] and 
magnetohydrodynamics [3]. In the third type, numerical simulations are carried out to observe
how matter is being deflected from the equatorial plane towards the axis [4,5,6,7,8]. From the 
analytical front, although the wind type and the accretion type solutions come out from the same set
of governing equations, there were no attempt to find connections among them. As a result, the 
computation of the outflow rate directly from the inflow parameters has always been impossible.
Our work, {\it for the first time}, quantitatively connects the topologies of the inflow and the
outflow to give a self consistent model for computing the mass loss rate from compact objects.\\
The massoutflow rates from the ordinary stellar bodies have been calculated very accurately from 
the stellar luminosity because theory of radiatively driven winds seems to be well understood [9].
The most fundamental difference between the mass loss from normal stellar objects and compact objects
is that the stellar bodies have their own atmosphere from which the outflowing
mass is ejected out. On the other hand, compact objects, such as a black hole or a nutron star,
do not have atmosphere of their own and wind must be generated from the inflow only. Given that the
accretion disks surrounding them are sufficiently hot (near the inner edge) to be ionized in general, 
similar method as employed in stellar atmosphere could be applicable to the compact objects. Our approach in this
work is precisely this. We first 
determine the properties of the rotating inflow and outflow and
identify solutions to connect them. In this manner we self-consistently determine the mass outflow rates.
In case of quasi-spherical accretion with almost zero angular momentum, the accretion disk does
not form. There the pressure of pair plasma creates the virtual boundary layer wich mimics the 
stellar atmosphere for our purpose.\\
Before we present our results, we describe basic properties of
the rotating inflow and outflow. A rotating inflow with a specific angular momentum $\lambda$
entering into a black hole  will have angular momentum $\lambda \sim $  constant
close to the black hole
for any moderate viscous stress. This is because the
viscous time scale is generally much longer compared to the
infall time scale. Almost constant angular momentum
produces a very strong centrifugal force $\lambda^2/r^3$ which
increases much faster compared to the gravitational force
$\sim GM/r^2$ and becomes comparable at around $r\sim l^2/GM$,
or, $x_{cb}\sim 2\lambda^2$ where $x$ and $\lambda$ are $r$ and $l$,
written in units of $R_g=2GM/c^2$ and $R_g c = 2GM/c$ respectively.
The subscript $cb$ under $x$ stands for the centrifugal barrier. Here,
(actually a little farther out, due to thermal pressure) matter starts
piling up and produces the centrifugal pressure supported boundary layer
(CENBOL). Farther close to the black hole, the gravity always wins
and matter enters the horizon supersonically after passing
through a sonic point. CENBOL may or may not have a sharp boundary,
depending on whether standing shocks form or not. Generally speaking,
in a polytropic flow, if the polytropic index $\gamma > 1.5$, then
shocks do not form and if $\gamma <1.5$, only a region of the parameter
space forms the shock [12]. In any case, the CENBOL forms. In this region
the flow becomes hotter and denser and for all practical purposes
behaves as the stellar atmosphere so far as the formation of
outflows are concerned. Inflows on neutron stars behave similarly,
except that the `hard-surface' inner boundary condition dictates that the
flow remain subsonic between the CENBOL and the surface rather than becoming
supersonic as in the case of a black hole. In case where the shock does not form,
regions around pressure maximum achieved just outside the inner sonic
point would also drive the flow outwards.
Outflow rates from accretion disks around  black holes and neutron
stars must be related to the properties of CENBOL which in turn,
depend on the inflow parameters. Subsonic outflows originating from CENBOL
would pass through sonic points and reach far distances as in wind
solutions. The following 3d cartoon diagram shows the schematic geometry
of the disk-jet system. The arrows show the axis of the 
whirling jet, D(K) stands for the Keplarian part of the disk
and D(SK) stands for the subkeplarian part. CENBOL forms 
somewhere inside the D(SK) and J stands for the jet structure.\\
\begin{center}
  \leavevmode
  \epsfbox{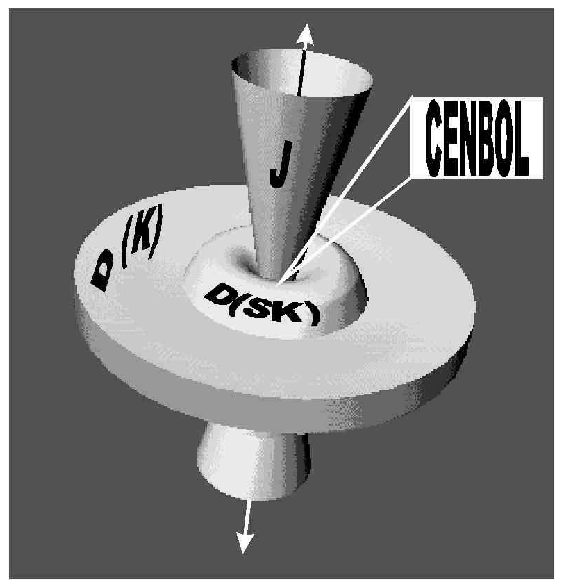}
\end{center}
\begin{center}
\underline
{\bf Geometry of the disk-jet system}
\end{center}
\noindent
Considering the inflow to be polytropic, we explore
both the polytropic and the isothermal outflow.
After defining the mass outflow rate as
$$
\frac{{\dot M}_{out}}{{\dot M}_{in}} = R_{\dot m} ,
$$
We calculate this rate as a function of the inflow parameters, such
as specific  energy and angular momentum, accretion rate, polytropic index etc.
A detail report of this work is presented elsewhere [10].
\noindent
In the absence of accretion disks (inflow with almost zero angular momentum), the freely
falling matter onto compact objects 
becomes supersonic 
after crossing the sonic point
and  may produce a standing collisionless shock due to the
plasma instabilities and the nonlinearity introduced in the flow due to density perturbation 
[11]. Once the shock is formed, the relativistic protons provide the 
necessary pressure to support the shock. This standoff shock layer then acts as the 
CENBOL to produce outflows for the quasi-spherical bondi type models of AGNs.\\
The plan of this paper is what follows: In the next section, we describe our model 
(when disks are formed) and present the governing equations for the inflow and
outflow along with the simultanious solution procedure. In \S 2, we present the
results of our computation. In \S 3, we will discuss about the mass outflow from
accretion with almost zero angular momentum (bondi type accretion) and present
the schematic results. Finally, in \S 4, we draw our conclusions.\\

\section{Model Description, Governing Equations and the solution Procedure}

%\subsection{Inflow Model}
We consider thin, axisymmetric polytropic inflows in vertical equilibrium 
(otherwise known as 1.5 dimensional flow). We 
ignore the self-gravity of the flow and viscosity is assumed to be significant only
at the shock so that entropy is generated. We do the calculations using Paczy\'nski-Wiita 
potential which mimics surroundings of the Schwarzschild black hole. 
The equations (in dimensionless units) governing the inflow are:\\
$$
{\cal E}=\frac{{u_e}^2}{2} +n{{a_e}^2}+
\frac{{\lambda}^2}{2{r^2}}-\frac{1}{2(r-1)} .
\eqno{(1)}
$$
$$
{{\dot M}_{in}}={u_e}{{\rho}_e}r{h_e}(r),
\eqno{(2)}
$$
(For detail, see [12] )
The equations governing the polytropic outflow are
$$
{\cal E}=\frac{\vartheta^2}{2}+{n^\prime}{{a_e}^2}+\frac{{\lambda}^2}{2{{r_m}^2(
r)}}
-\frac{1}{2(r-1)}
\eqno{(3)}
$$
$$
{{\dot M}_{out}}={\rho}{\vartheta}{\cal A}(r).
\eqno{(4)}
$$
Where ${r_m}$ is the mean axial distance of the flow and ${{\cal A}(r)}$
is the cross sectional area through which mass is flowing out. (For detalil,
see, [10]) $\gamma$ of the outflow was taken to be smaller than that of the inflow
because of momentum deposition effects.
The outflow angular momentum $\lambda$ is chosen to be the same as in the
inflow, i.e., no viscous dissipation is assumed to be present in the inner region of the
flow close to a black hole. Considering that viscous time scales are longer compared to 
the inflow time scale, it may be a good  assumption in the disk, but it may not be
a very strong assumption for the outflows which are slow prior to the acceleration and
are therefore, prone to viscous transport of angular momentum.  
Detailed study of the outflow rates in presence of viscosity 
and magnetic field is in progress and would be presented elsewhere.
The Isothermal outflow is governed by the following equations
$$
\frac{{\vartheta_{iso}}^2}{2}+{{C_s}^2}ln{\rho}+\frac{\lambda^2}{2{{r_m}(r)}^2}
-\frac{1}{2(r-1)}={\rm Constant}
\eqno{(5)}
$$
$$
{{\dot{M}}_{out}}=\rho {\vartheta_{iso}}{{\cal A}(r)}.
\eqno{(6)}
$$
Here, the area function remains the same above. A subscript {\it  iso} of 
velocity ${\vartheta}$ is kept to distinguish from the velocity in 
the polytropic case. This is to indicate the 
velocities are measured here using completely different assumptions. For details, see [10].

In both the models of the outflow, we assume that the flow is primarily radial. Thus the
$\theta$-component of the velocity is ignored ($\vartheta_\theta  << \vartheta$).

\subsection{Procedure to solve for disks and outflows simultaneously}

For polytropic outflows, we solve equations (1-4) simultaneously using numerical
techniques (for detail, see, [10]).
In this case the specific energy ${\cal E}$ is assumed to  
remain fixed throughout the flow trajectory as it moves from the disk to the jet. 
At the shock, entropy is generated
and hence the outflow is of higher entropy for the same specific energy.

A supply of parameters ${\cal E}$, $\lambda$, $\gamma$ and $\gamma_{o}$ makes
a self-consistent computation of $R_{\dot m}$ possible when the shock is present. 
In the case where the shocks do not form, the procedure is a bit different. 
It is assumed that the maximum amount of matter comes out from the 
place of the disk where the thermal pressure of the inflow attains its maximum 
and the outflow is assumed to have the same quasi-conical shape with annular
cross-section ${\cal A} (r)$ between the funnel wall and the centrifugal barrier as already defined.
For this case, the compression ratio of the gas at the pressure maximum between the inflow and outflow
$R_{comp}$ is supplied as a free parameter, since it may be otherwise very difficult to 
compute satisfactorily. In the presence of shocks, such problems do not arise as the
compression ratio is obtained self-consistently. For isothermal outflow, 
it is assumed that the outflow has exactly the {\it same} temperature as that of the
post-shock flow, but the energy is not conserved as matter goes from disk to the 
wind. The polytropic index of the inflow can vary but that of the outflow is
always unity. The other assumptions and logical steps are exactly same
as those of the case where the outflow is polytropic. Here we solve 
equations (1-2) and (5-6) simultaneously using numerical technique to get 
results. (For details, see, [10]). 

\section{Results}

\subsection{Polytropic outflow coming from the post-shock accretion disk}
Figure 1 shows a typical solution which combines the accretion and the outflow. The input parameters
are ${\cal E}=0.00689$, ${\lambda=1.65}$ and $\gamma=4/3$ corresponding to 
relativistic inflow. The solid curve with an arrow represents 
the pre-shock region of the inflow and the long-dashed curve represents the post-shock 
inflow which
\begin{figure}
\vbox{
\vskip 0.0cm
\hskip 0.0cm
\centerline{
\psfig{figure=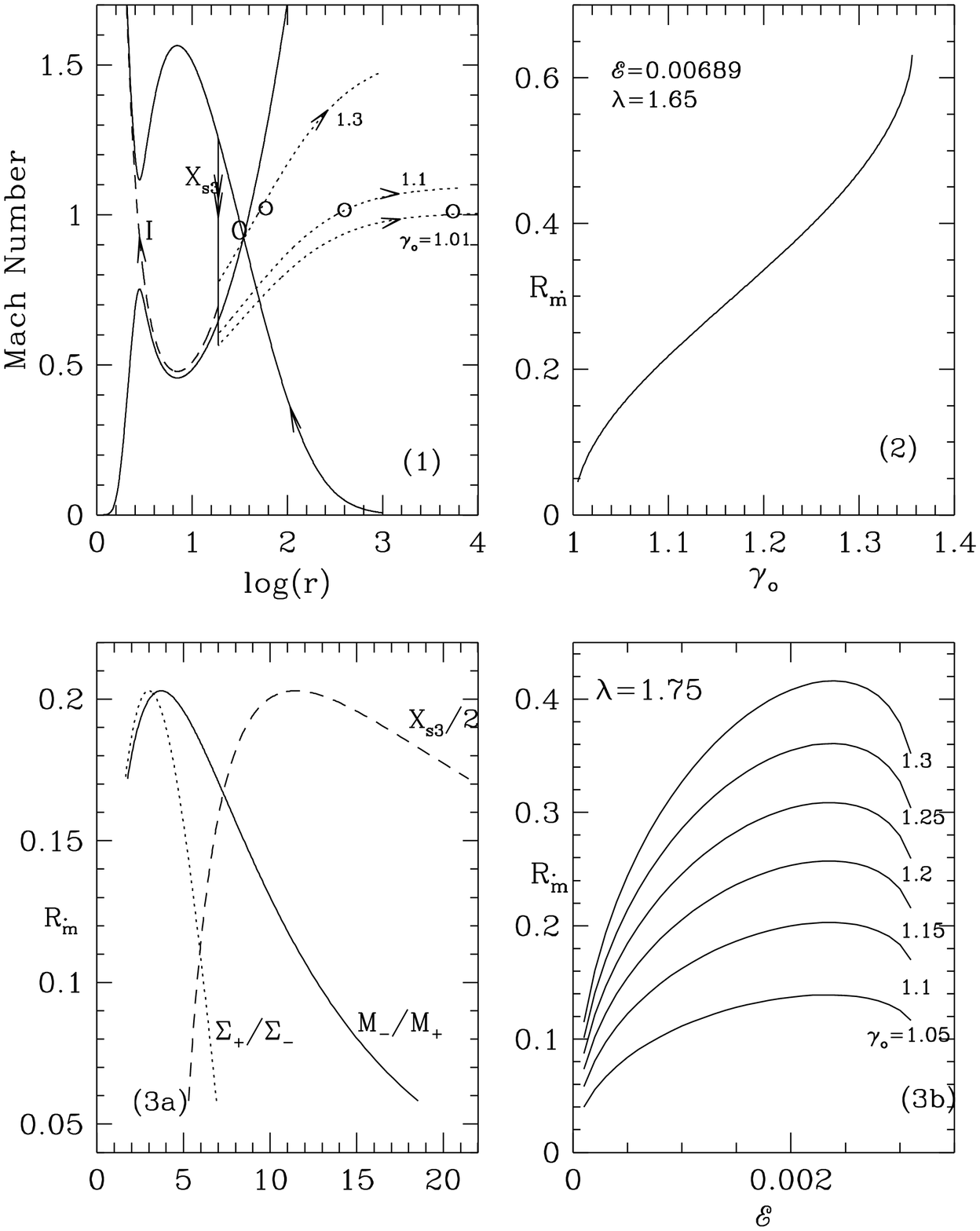,height=10truecm,width=10truecm,angle=0}}}
\vskip 0.5cm
\noindent{\small {\bf Fig. 1-3}: Mach number of the flow is plotted against logarithmic radial
distance both for the inflow and outflow (Fig. 1). The ratio of mass outflow rate and mass inflo
w rate
is plotted against the polytropic index of the outflow (Fig. 2). The same ratio is plotted again
st
the shock strength, shock location and the ratio of the integrated density  (Fig. 3a) and specif
ic
energy ${\cal E}$ and polytropic index of the outgoing flow $\gamma_o$ (Fig. 3b).
See text for details.}

\end{figure}
\noindent
enters the black hole after passing through the inner sonic point (I).
The solid vertical line at $X_{s3}$ (in the notation of [12]) 
with double arrow represents the shock transition.
Three dotted curves represent three outflow solutions for the parameters 
$\gamma_{o}=1.3$ (top), $1.1$ (middle) and $1.03$ (bottom). The outflow
branches shown pass through the corresponding sonic points. It is
evident from the figure that the outflow  moves along solution curves which are completely different from that
of the `wind solution' of the inflow which passes through the outer sonic point `O'.
The mass loss ratio $R_{\dot m}$ in these cases are $0.47$, $0.22$ and $0.06$ 
respectively. Figure 2 shows the ratio $R_{\dot m}$ as $\gamma_{o}$ is varied. Only the 
range of $\gamma_{o}$ for which the shock-solution is present is shown here. 
In Fig. 3a we show the variation of the ratio $R_{\dot m}$ 
of the mass outflow rate inflow rate as a function of 
the shock-strength (solid) $M_-/M_+$ (Here, $M_-$ and $M_+$ 
are the Mach numbers of the pre- and post-shock flows respectively.),
the compression ratio (dotted) $\Sigma_+/\Sigma_-$ (Here, $\Sigma_-$ and $\Sigma_+$ 
are the vertically integrated matter densities in the pre- and post- shock
\begin{figure}
\vbox{
\vskip 0.0cm
\hskip 0.0cm
\centerline{
\psfig{figure=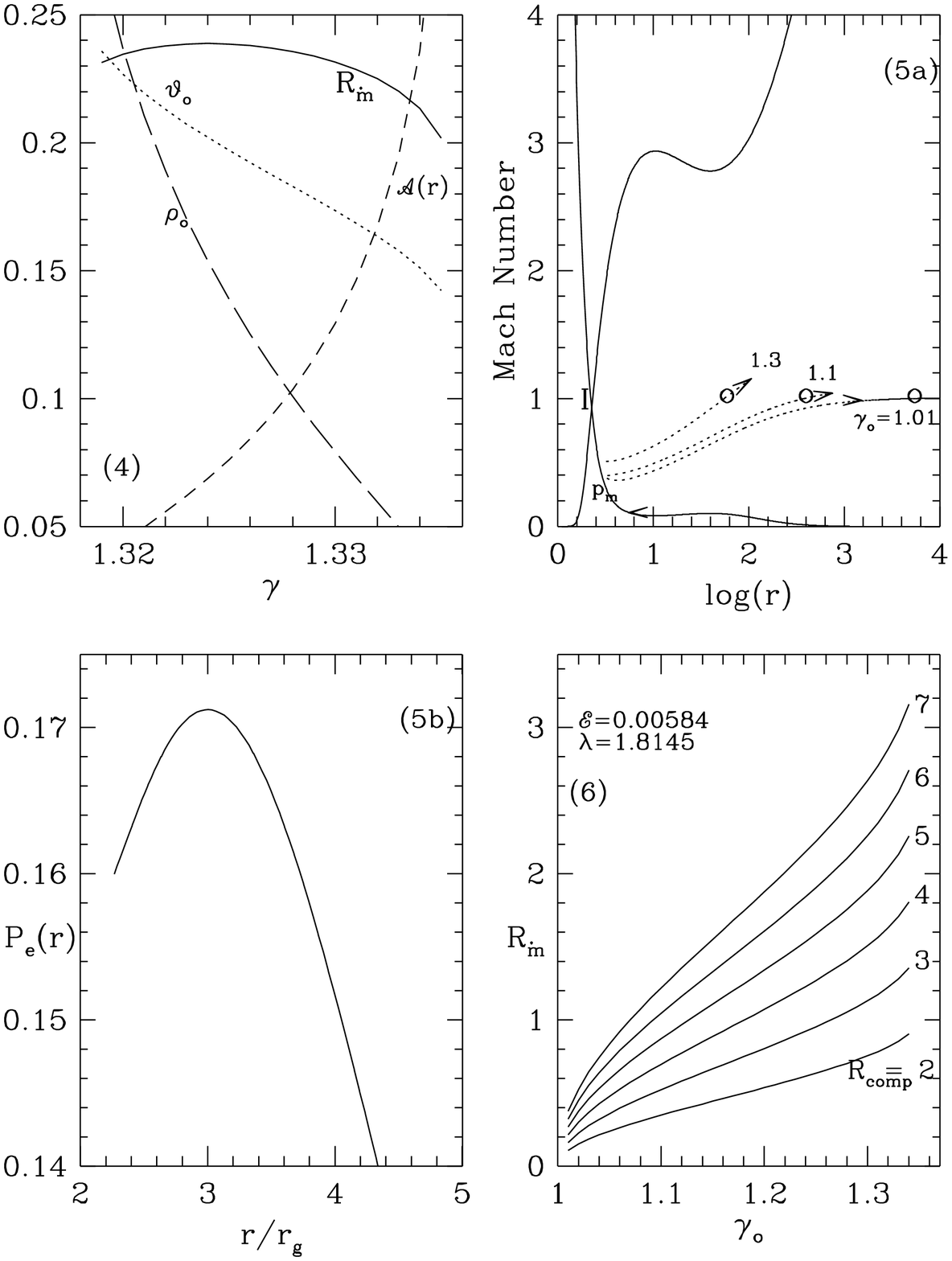,height=10truecm,width=10truecm,angle=0}}}
\vskip 0.5cm
\noindent{\small {\bf Fig. 4-6}: Variation of velocity, density, cross sectional
area and the rate ratio as a function of the polytropic index of the inflow (Fig. 4).
Variation of the Mach number for inflow and outflow when shocks are not present (Fig. 5a).
Thermal pressure variation as a function of the radial distance $r/r_g$ showing
a distinct maximum (Fig. 5b). Variation of $R_{\dot m}$ when both the
compression ratio at the pressure maxima and plytropic index of the outflow
are changed (Fig. 6). See text for details.}
\end{figure}
\noindent
flows respectively), and the stable shock location (dashed) $X_{s3}$. 
Other parameters are $\lambda=1.75$ and $\gamma_{o} = 1.1$. 
Note that the ratio $R_{\dot m}$ does not peak near the strongest 
shocks! Shocks are stronger when they are located closer to the black 
hole, i.e., for smaller energies. 
In Fig. 3b where $R_{\dot m}$ is plotted 
as a function of the specific energy ${\cal E}$ (along x-axis) 
and $\gamma_{o}$ (marked on each curve). Specific angular momentum 
is chosen to be $\lambda=1.75$ as before. 
To have a better insight of the behavior of the outflow we plot in Fig. 4 $R_{\dot m}$
as a function of the polytropic index of the incoming flow $\gamma$. 
The range of $\gamma$ shown is the range for which shock forms in the flow. We also plot the
variation of velocity $\vartheta_o$, density $\rho_o$ and area ${\cal A}(r)$ of the outflow
at the location where the outflow leaves the disk.
These quantities are scaled from the corresponding dimensionless units as $\vartheta_o \rightarrow 
2 \times 10^4 \vartheta_o -558$, $\rho_o \rightarrow 10^{22} \rho_o$ and ${\cal A} 
\rightarrow 0.0005 {\cal A} $ respectively in order to bring them in the
same scale. The non-monotonic nature of the variation of $R_{\dot m}$ with $\gamma$ is observed.
\subsection{Polytropic outflow coming from the region of the maximum pressure}
In this case, the inflow parameters are chosen from such a region of parameter space
so that the shocks do not form (see [12]). Here, the inflow passes through 
the inner sonic point only. The outflow is assumed to be coming 
out from the regions where the polytropic inflow has maximum pressure. 
Figure 5a shows a typical solution. The arrowed solid curve shows the inflow 
and the dotted arrowed curves show the outflows 
for $\gamma_o=1.3$ (top), $1.1$ (middle) and $1.01$ (bottom). 
The ratio $R_{\dot m}$ in these cases 
is given by $0.66$, $0.30$ and $0.09$ respectively. 
The specific energy and angular momentum are chosen to be ${\cal E}=0.00584$
and $\lambda=1.8145$ respectively.
The pressure maximum occurs outside the inner sonic point at $r_m$ when 
the flow is still subsonic. Figure 5b shows the variation of 
thermal pressure of the flow with radial distance. The peak is clearly visible. 
Figure 6 shows the 
ratio $R_{\dot m}$ as a function of $\gamma_{o}$ for various choices of the
compression ratio $R_{comp}$ 
of the outflowing gas at the pressure maximum: $R_{comp}=2$ for the
bottom curve and $7$ for the top curve. Note that flows with 
highest compression ratios
produce highest outflow rates, evacuating the disk which is responsible
for the quiescent states in X-ray Novae systems and also in some systems with massive black holes
(e.g., our own galactic centre?). 
The location of maximum pressure being close to the black hole, it  may be 
very difficult to generate the outflow from this region. Thus, it is expected that the
ratio $R_{\dot m}$ would be larger when the maximum pressure is located farther out.
This is exactly what we see in Fig. 7, where we plot $R_{\dot m}$ against the location of
the pressure maximum (solid curve). Secondly, if our guess that the outflow rate could be related to the
pressure is correct, then the rate should increase as the pressure at the maximum rises.
That's also what we observe in Fig. 7. We plot $R_{\dot m}$ as a function of the 
actual pressure at the pressure maximum (dotted curve). The mass loss is found to be a strongly
correlated with the thermal pressure. 
Here we have multiplied non-dimensional thermal 
pressure by $1.5 \times 10^{24}$ in order to bring them in the same scale.
\subsection{Isothermal outflow coming from the post-shock accretion disk}
Here the temperature of the outflow
is obtained from the proton temperature of the advective region of the disk. The proton temperature
is obtained using the Comptonization, bremsstrahlung, inverse bremsstrahlung and Coulomb processes. 
[13]. Figure 8 shows the effective
proton temperature and the electron temperature of the post-shock advective region as a function of the
accretion rate (in logarithmic scale)  of the Keplerian component of the disk. 
In Fig. 9a, we show the ratio $R_{\dot m}$ as a function of the Eddington rate of the
incoming flow for a range of the specific angular momentum. In the low luminosity
objects the ratio
\begin{figure}
\vbox{
\vskip 0.0cm
\hskip 0.0cm
\centerline{
\psfig{figure=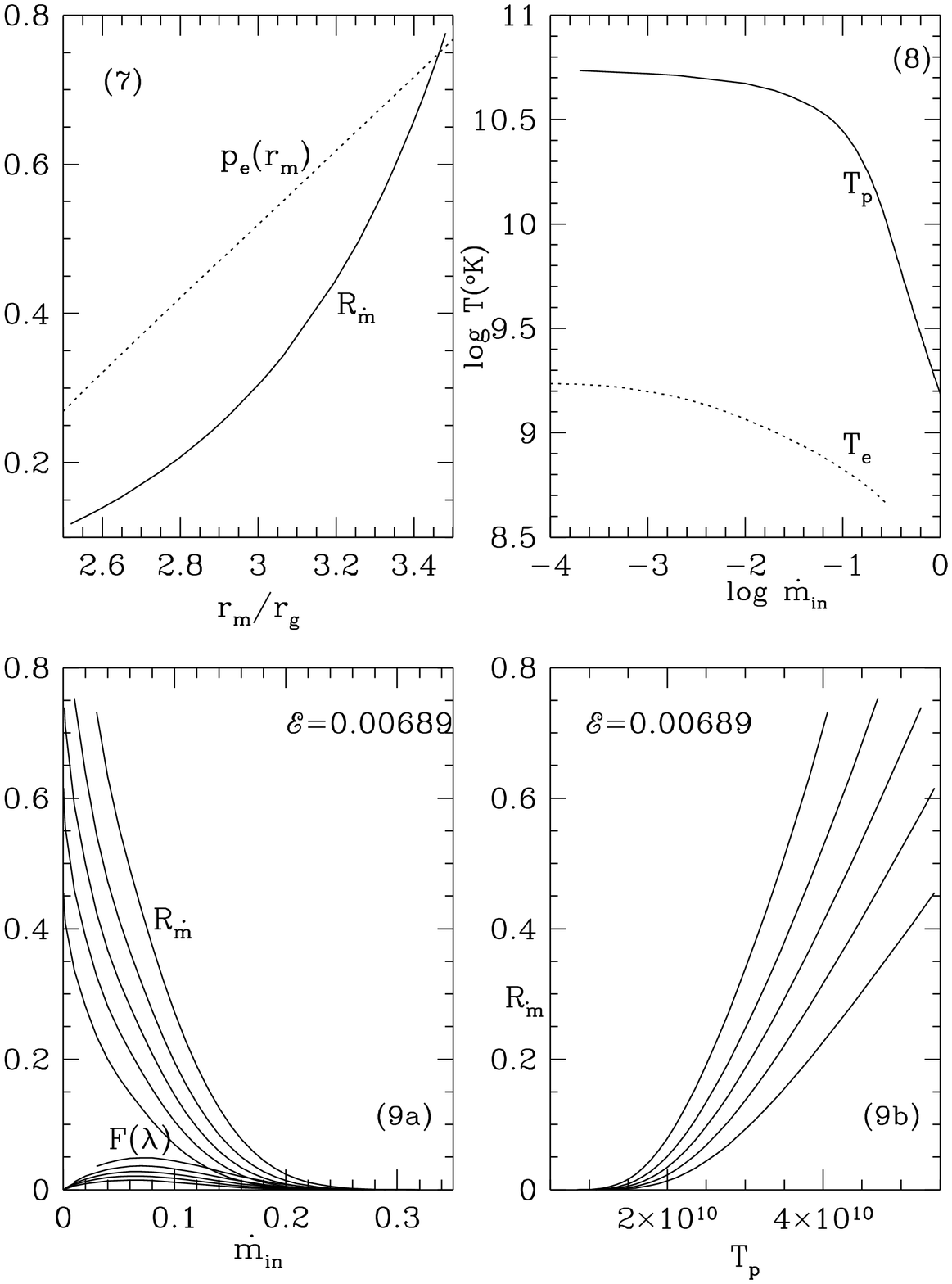,height=10truecm,width=10truecm,angle=0}}}
\vskip 0.5cm
\noindent{\small {\bf Fig. 7-9}: Variation of the maximum pressure and $R_{\dot m}$
with the location where the pressure maxima occur (Fig. 7). Proton and electron
temperatures in the advective region as a function of the inflow disk accretion rate
${\dot m}_{in}$ (Fig. 8). Variation of the $R_{\dot m}$ and angular momentum
flux $F(\lambda)$ as a function of the accretion rate of the inflow (Fig. 9a). Variation of
$R_{\dot m}$ with proton temperature $T_p$ (Fig. 9b). See text for details.}
\end{figure}
\noindent
is larger. Angular momentum is varied from $\lambda=1.63$ (top curve) 
to $1.65$ (bottom curve). An interval of $\lambda=0.005$ was used.
The ratio is very sensitive to the angular momentum since it changes the
shock location rapidly and therefore changes the post-shock temperature very much. 
We also plot the outflux of angular momentum $F ({\lambda})=\lambda {\dot m}_{in} R_{\dot m}$
which has a maximum at intermediate accretion rates. In dimensional units, these
quantities represent significant fractions of angular momentum  of the entire disk
and therefore the rotating outflow can help accretion processes. Curves are drawn for different $\lambda$
as above. In Fig. 9b, we plot the  variation of the ratio directly with the proton temperature of the
advecting region. The outflow is clearly thermally driven. Hotter flow 
produces more winds as is expected. The angular momentum associated
with each curve is same as before.
\subsection{Isothermal outflow coming from the region of the maximum pressure}
This case produces very similar result as in the above case, except that like Section 2.2 the
outflow rate becomes more than a hundred percent of the inflow rate when the proton temperature is very high. 
This phenomenon may be responsible for producing quiescent states in some black hole
candidates.
\section{Poytropic outflow from quasi-spherical accretion}
\noindent
As already been mentioned in the introduction, for a quasi-spherical Bondi-type
accretion onto a compact object (namely, on a massive black hole), a steady
state situation can be developed where a standing collisionless shock
may form due to the plasma instabilities and for nonlinearyity introduced
by small density purturbation. We consider cold inflow ( ${{\cal E} \sim  0.001}$)
with moderate or high value of accretion rate. Thermal particles freely falling
toward black hole are assumed to be shock accelerated via first order Fermi
acceleration producing relativistic protons. Those relativistic protons
usually scatter several times before being captured by the black hole.
These energyzed particles, in turn, provide sufficiently outward 
pressure to support a standing,
collisionless shock. A fraction of the energy flux of infalling matter
is assumed to be converted into radiation at the shock standoff 
distance through hadronic collision and mesonic decay. That fraction is 
radiated away to support the development and maintanence of a standing,
collisionless shock at a given Schwarzschild radius. [11] \\
The fraction of energy converted, the shock compression ratio 
${R_{comp}}$, along with the ratio of post shock relativistic
hadronic pressure to infalling infalling ram pressure at a given shock location
are obtained from the steady state shock solution of Eichler [14]
and Ellision and Eichler [15].
The shock location as a function of the specific energy ${\cal E}$ of the 
infalling matter and accretion rate is then self consistently obtained
using the above mentioned quantities.\\
We consider polytropic inflow. The outflow is also
assumed to be polytropic except the fact that ${{\gamma}_{outflow}}$
is assumed to be
less than the ${{\gamma}_{inflow}}$ reason of which has already been
discussed in \S 1. 
As a fraction of the energy of infalling material is converted into
radiation, energy flux of the wind is somewhat less than that of the accretion
but is kept constant througout the outflow. Below we present a priliminary report
of the result. Detail calculation is presented elsewhere [16].
\subsection{Priliminary report of the results}
We solve the inflow and outflow equations 
self consistently as was described in \S 1.1 except the fact that now there
will be no angular momentum related term in the energy expression and
as the flow is quasi-connical, the expressions for mass flux rates are now
$$
{\dot M}_{(in/out)}={\rho}u{{r^2}_{shock}}{{\Theta}_{(in/out)}}
$$
The value of ${R_{\dot m}}$ is distinguishably small compared to the cases
previously discussed (i,e, for the rotating flows). This is because matter is
ejected out due to the presure of the relativistic particles generated 
at the shock location which is less
enough in comparison to that originated for the presence of angular
momentum. In general mass loss rate increases with the energy of the infall
for fixed accretion rate. When energy is varied, massloss rate is 
anti-correlated
with the shock location. This is expected because higher energy gives the 
lower value of shock location in our model and the closer the shock forms
to the black hole, the greater will be the amount of gravitational
potential available to be put onto relativistic hadrons to apply pressure
for the outflow to take place. If the energy is kept fixed and accretion rate is varied
it is observed that the shock location increases with increasing accretion rate as 
expected from the functional form of the formula for calculating the 
shock standoff distance. Here the mass loss rate is correlated 
with the shock location and
accretion rate. This is because once the energy is fixed, the fraction of it which
is converted into radiation is also fixed implying the fact that lower is the shock 
location, harder is the job to produce the outflow because of the inward pull of gravity
strength of which increases as the shock location decreases (for detail, see [16]).\\
Variation of ${R_{\dot m}}$ with the compression ratio ${R_{comp}}$ 
at the shock 
location follows more or less the same trend as was manifested for the flow
with angular momentum (Fig - 3a) except that here the peak is not that much distinct,
and, in general, increases with the ${{\gamma}_{outflow}}$ as was seen in previous
case (Fig - 2). See [16] for detail calculations and figures.\\

\section {Concluding remarks :}
In this paper, we have computed the mass outflow rate from the matter accreting
onto galactic and extra-galactic black holes. Since the general physics
of advective flows are similar around a neutron star, we believe that the
conclusions may remain roughly similar provided the shock forms,
although the boundary layer 
of the neutron star, where half of the binding energy could be released,
 may be more luminous than that of a black hole and may thus
affect the outflow rate.\\
\noindent  
From numerical simulation using SPH code, it was found that the outflow rate
could be as high as 10 - 20 per cent [6] which we get for moderate accretion rate
and for the low value of ${\gamma_{out}}$. Using TVD code [8], 10 - 15 per cent of the steady
outflow is seen and occasionally, even 150 per cent of the inflow is found to be driven away.
Our result shows that similar high outflow rate is also possible, especially for low 
luminosities. Simulation for radiation dominated flows showed ${{R_{\dot m}} \sim 0.004}$ 
[5], which also agrees with our results when we consider high accretion rate (see Fig. 9a).
So it seems that the analytical results of our work are in good agreement with 
numerical simulation work.
Observationally,
the exact value of outflow rate from a real system is very 
difficuilt to obtain as it depends on
too many uncertainties, such as filling factors and projection effects etc. In any case, with a
general knowledge of the outflow rate, we can proceed to estimate several important quantities.
For example, it had been argued that  the composition of the disk changes due to nucleosynthesis
in accretion disks around black holes and these modified isotopes are deposited in the
surroundings by outflows from the disks (Hogan \& Applegate 1987; Mukhopadhyay \& Chakrabarti, 1998
and references therein). Similarly, it is argued that outflows deposit magnetic flux tubes from accretion disks
into  the surroundings (Daly \& Loeb, 1990). Thus a knowledge of outflows are essential in understanding
varied physical phenomena in galactic environments.\\
The basic conclusions of this paper (for flows with angular momentum)
 are the followings:\\
 \noindent a) It is possible that most of the outflows are coming from the centrifugally
supported boundary layer (CENBOL) of the accretion disks.\\
\noindent b) The outflow rate generally increases with the proton temperature of CENBOL. In other
words, winds are, at least partially, thermally driven.
This is reflected more strongly when the outflow is isothermal.\\
\noindent c) Even though specific angular momentum of the flow increases the size of the CENBOL,
and one would have expected a higher mass flux  in the wind, we find that the rate of the
outflow is actually anti-correlated with the $\lambda$ of the inflow. On the other hand, presenc
e of
significant viscosity in CENBOL may reduce angular momentum of the outflow.  When this
is taken into account, we find that the rate of the outflow is correlated with
$\lambda$ of the outflow. This suggests that the outflow is partially centrifugally driven as well.\\
\noindent d) The ratio $R_{\dot m}$ is generally anti-correlated with the inflow accretion rate.
 That is,
disks of lower luminosity would produce higher $R_{\dot m}$.\\
\noindent e) Generally speaking, supersonic region of the inflow do not have pressure maxima. Thus, 
outflows emerge from the subsonic region of the inflow, whether the shock actually forms or not.
\\
\noindent
An interesting situation arises when the polytropic index of the outflow is large
and the compression ratio of the flow is also very high. In this case, the flow virtually
bounces back as the winds and the outflow rate can be temporarily larger compared with the
inflow rate, thereby evacuating the disk. In this range of parameters, most, if not all,
of our assumptions breakdown completely because the situation becomes inherently time-dependent.
It is possible that some of the black hole systems, including that in our own galactic centre,
may have undergone such evacuation phase in the past and gone into quiescent phase.\\
So far, we made the computations around a Schwarzschild black hole.
The mass outflow rates for kerr black holes
are being studied and the results would be reported elsewhere [17].
We made a few assumptions, some of which may be questionable.
Nevertheless, we believe that our calculation
is sufficiently illustrative and gives a direction which can be followed in the future.\\
\begin{center}
{\large\bf REFERENCES}
\end{center}
\begin{enumerate}
\item {Blandford, R. D. $\&$ Payne, D. 1982, MNRAS, 199, 883}
\item {Chakrabarti, S. K. 1986, Apj, 303, 582}
\item {Chakrabarti, S. K. $\&$ Bhaskaran, P. 1992, 255, 255}
\item {Hawley, J.W., Smarr, L. $\&$  Wilson, J. 1984, Apj, 297, 296.}
\item {Eggum, G. E., Coroniti, F. V., Katz, J. I. 1985, Apj, 298, L41}
\item {Molteni, D., Lanzafame, G. $\&$  Chakrabarti, S. K. 1994 Apj, 425, 161}
\item {Nobuta, K., Hanawa, T. 1997 IAUS..184E.245N}
\item {Ryu, D., Chakrabarti, S. K., $\&$  Molteni, D. 1997, Apj, 474, 378}
\item {Castor, J. I., Abott, D. C. $\&$ Klein, R. I. 1975, Apj, 195, 157}
\item {Das, T., K. $\&$ Chakrabarti, S., K. 1998 Apj, (submitted)}
\item {Kazanas, D. $\&$ Ellison, D. C. 1986, Apj 304, 178.}
\item {Chakrabarti, S. K. 1989, Apj, 347, 365}
\item {Chakrabarti, S. K. 1997, Apj, 484, 313}
\item {Eichler, D., 1984, Apj., {\bf 277}, 429}
\item {Ellision, D. C., $\&$ Eichler, D. 1984, Apj, {\bf 286}, 691}
\item {Das, T., K. 1998 (submitted)}
\item {Das, T., K. 1998, in preparation}
\item {Chakrabarti, S. K. 1997, Apj, submitted}
\end{enumerate}
\end{document}